\documentclass[twocolumn,showpacs,preprintnumbers,amsmath,amssymb]{revtex4}

\usepackage{graphicx}
\usepackage{dcolumn}
\usepackage{bm}

\usepackage{amssymb}
\usepackage{amsmath}

\begin{document}


\title{Cosmic microwave background radiation temperature in a dissipative universe}

\author{Nobuyoshi {\sc Komatsu}$^{1}$}  \altaffiliation{E-mail: komatsu@se.kanazawa-u.ac.jp} 
\author{Shigeo     {\sc Kimura}$^{2}$}

\affiliation{$^{1}$Department of Mechanical Systems Engineering, Kanazawa University, 
                          Kakuma-machi, Kanazawa, Ishikawa 920-1192, Japan \\
                $^{2}$The Institute of Nature and Environmental Technology, Kanazawa University, 
                          Kakuma-machi, Kanazawa, Ishikawa 920-1192, Japan}%
\date{\today}

\begin{abstract}
The relationship between the cosmic microwave background radiation temperature and the redshift, i.e., the $T$--$z$ relation, is examined in a phenomenological dissipative model. 
The model contains two constant terms, as if a nonzero cosmological constant $\Lambda$ and a dissipative process are operative in a homogeneous, isotropic, and spatially flat universe.
The $T$--$z$ relation is derived from a general radiative temperature law, as appropriate for describing nonequilibrium states in a creation of cold dark matter (CCDM) model. 
Using this relation, the radiation temperature in the late universe is calculated as a function of a dissipation rate ranging from $\tilde{\mu} =0$, corresponding to a nondissipative $\Lambda$CDM model, to $\tilde{\mu} =1$, corresponding to a fully dissipative CCDM model. 
The $T$--$z$ relation for $\tilde{\mu} =0$ is linear for standard cosmology and is consistent with observations.
However, with increasing dissipation rate $\tilde{\mu}$, the radiation temperature gradually deviates from a linear law because the effective equation-of-state parameter varies with time.
When the background evolution of the universe agrees with a fine-tuned pure $\Lambda$CDM model, the $T$--$z$ relation for low $\tilde{\mu}$ matches observations, whereas the $T$--$z$ relation for high $\tilde{\mu}$ does not.
Previous work also found that a weakly dissipative model accords with measurements of a growth rate for clustering related to structure formations.
These results imply that low dissipation is likely for the universe.
The weakly dissipative model should be further constrained by recent observations.

\end{abstract}

\maketitle

\section{Introduction}

The blackbody radiation temperature of the cosmic microwave background is $T_{0} = 2.725 \pm 0.002$ K at the present time \cite{Mather_1999}.
This temperature is regarded as the evidence for a hot Big Bang.
That hypothesis is supported by measurements of the radiation temperature--redshift $T$--$z$  relation \cite{Songaila_1994,Lu_1996,Ge_1997,Srianand_2000,Molaro_2002,Cui_2005,Srianand_2008,Luzzi_2009,Noterdaeme_2010,Noterdaeme_2011,Martino_2015,Luzzi_2015}.
The observations are consistent with a linear law, $T = T_{0} (1+z)$, as can be derived from standard cosmology, such as $\Lambda$CDM (lambda cold dark matter) models. 
However, the $\Lambda$CDM model has several theoretical difficulties \cite{Weinberg1989,Weinberg1,Roy1,Bamba1_Miao1}, although it can elegantly explain the accelerated expansion of the late universe \cite{PERL1998ab,Riess1998_2004,Riess2007SN1,Suzuki_2012,Planck2013,Planck2015}. 
To explain the acceleration, various models have been suggested, 
such as $\Lambda (t)$CDM which assumes a time-varying cosmological term \cite{Freese-Lima,Sola_2009,Sola_2011a,Sola_2013b,Sola_2015c}, 
bulk viscous models which assume a bulk viscosity for the cosmological fluid \cite{Weinberg0,Murphy1,Barrow1986-Lima1988,Zimdahl1996-Nojiri2011}, 
and CCDM which assumes the creation of cold dark matter \cite{Lima_2000,Lima_1992b,Lima_1992e,Lima_2014b,Harko_2014,Harko_2013,Zimdahl_1993,Lima-Others1996-2008,Lima2010,Lima2010b,Lima2011,Lima2012,Jesus2014,Ramos_2014-2014b}. 

The bulk viscous and CCDM models presume the existence of irreversible entropy in a homogeneous isotropic universe, unlike the $\Lambda$CDM and $\Lambda (t)$CDM models.
For example, in the CCDM model, irreversible entropy is generated from gravitationally induced particle creation in nonequilibrium thermodynamic states \cite{Prigogine_1988b,Prigogine1989}.
A possible equivalence of the bulk viscosity and matter creation dissipative mechanisms has been discussed in Ref.\ \cite{Lima_1992e}.
In addition, the connections between warm inflation \cite{Berera_1995-2014}, $\Lambda (t)$CDM models, and CCDM models have been debated in Ref.\ \cite{Connection}.
To examine the dissipative processes, a general radiative temperature law for adiabatic particle creation has been proposed by Lima \textit {et al.} \cite{Lima_2000,Lima_1992b,Lima_1992e,Lima_2014b}. 
That law has been examined from various viewpoints \cite{Lima_2014b,Harko_2014,Harko_2013}.
In particular, a simple $T$--$z$ relation of the form $T = T_{0} (1+z)^{1 - \beta}$ is frequently compared with observations, where $\beta$ is a constant parameter (cf. Refs.\ \cite{Lima_2000,Noterdaeme_2011}). 
The simple $T$--$z$ relation is obtained from a general radiative temperature law, if the effective equation-of-state parameter $w_{e}$ is constant \cite{Lima_2000,Lima_2014b}. 
However, in the CCDM model, $w_{e}$ varies during the evolution of the universe.
Thus, a time-varying $w_{e}$ needs to be considered when the radiation temperature is discussed in the CCDM model.
However, the radiation temperature in the CCDM model has not yet been quantitatively examined from this viewpoint.
It is important to do so in order to acquire a deeper understanding of the CCDM model.

In the CCDM model, a \textit{negative sound speed} \cite{Lima2011} and the existence of \textit{clustered matter} \cite{Ramos_2014-2014b} are necessary to properly describe the growth rate for clustering related to structure formations. 
Alternatively, a phenomenological dissipative model \cite{Koma7} has been proposed, in which the entropic force \cite{Easson12,Cai1-Costa1,Basilakos1,Sola_2014a,Koma4,Koma5,Koma6} is modified.
The model assumes constant terms that are equivalent to a nonzero cosmological constant $\Lambda$ and a dissipative process.
In previous work \cite{Koma7}, the dissipation rate was varied from $\tilde{\mu} =0$, corresponding to a nondissipative $\Lambda$CDM model, to $\tilde{\mu} =1$, corresponding to a fully dissipative CCDM model.
Low dissipation was found to correctly describe observations of structure formation.
The dissipation rate $\tilde{\mu}$ is expected to affect the $T$--$z$ relation, not only because the effective equation-of-state parameter $w_{e}$ depends on $\tilde{\mu}$, but also because $w_{e}$ varies during the evolution of the universe. 

This model makes it possible to examine a dissipative universe systematically, ranging from a nondissipative $\Lambda$CDM model to a fully dissipative CCDM model.
To clarify the properties of the radiation temperature in a dissipative universe, the $T$--$z$ relation can be examined.
A general radiative temperature law is applied to the dissipative model, to formulate the $T$--$z$ relation.
Based on this formulation, the radiation temperature in the late universe is calculated numerically as a function of the rate of dissipation.
The present study provides new insights and a unique approach for examining a dissipative universe.

The remainder of the article is organized as follows.
In Sec.\ \ref{Thermodynamics}, the general radiative temperature law for adiabatic particle creation is briefly reviewed.
In Sec.\ \ref{Dissipation model}, a phenomenological modified dissipative model is proposed.
In Sec.\ \ref{Temperature law for the dissipation model}, the temperature law is applied to the modified model, and the $T$--$z$ relation is formulated.
In Sec.\ \ref{Results}, the radiation temperature in a dissipative universe is examined.
Finally, in Sec.\ \ref{Conclusions}, the conclusions are presented.
(The modified dissipative model is different from dissipative particle dark matter models examined in Ref.\ \cite{Lisa_Foot}.)

\section{General radiative temperature law for adiabatic particle creation}
\label{Thermodynamics}
In this section, the general radiative temperature law for adiabatic particle creation is reviewed, following the work of Lima \textit{et al.} \cite{Lima_2000,Lima_1992e,Lima_1992b,Lima_2014b}.
A homogeneous, isotropic, and spatially flat universe is initially considered.
The line element given by the Friedmann--Robertson--Walker (FRW) metric \cite{Lima_1992e,Lima_2014b} is
\begin{equation}
  d s^{2} = c^{2} dt^{2} - a^{2}(t) ( dr^{2} + r^{2} d \theta^{2} + r^{2} \sin^{2} \theta d \phi^{2} )     
\label{eq:FRWmetric}
\end{equation}
where $c$ is the speed of light and $a(t)$ is the scale factor at time $t$.
The Hubble parameter $H$ is 
\begin{equation}
   H \equiv   \frac{ \dot{a}(t) } {a(t)}  .
\label{eq:Hubble}
\end{equation}

In what follows, nonequilibrium thermodynamic states of the cosmological fluid in a FRW background are considered. 
To this end, adiabatic particle creation is assumed \cite{Lima_1992e,Lima_1992b,Lima_2014b}.
The balance equations for the number of particles, entropy, and energy \cite{Lima_2014b} can then be written as
\begin{equation}
       \dot{n} + 3 H n   = n \Gamma   ,    
\label{eq:NonEquil_1}
\end{equation}
\begin{equation}
       \dot{s} + 3 H s   = s \Gamma   ,    
\label{eq:NonEquil_2}
\end{equation}
and
\begin{equation}
       \dot{\varepsilon} + 3 H ( \varepsilon + p + p_{c} )   = 0   ,
\label{eq:NonEquil_3}
\end{equation}
where $n$, $s$, $\varepsilon$, and $p$ are the particle number density, entropy density, energy density, and pressure, respectively. 
Here, $\Gamma$ and $p_{c}$ are the particle production rate and the dynamic creation pressure, respectively.
When $\Gamma =0$ and $p_{c}=0$, the three balance equations reduce to the conservation law for equilibrium states in a standard cosmology.
Equation \ (\ref{eq:NonEquil_1}) can be rewritten as $\dot{N} / N = \Gamma$, using the total number $ N \propto n a^3 $ of particles in the comoving volume \cite{Lima_2014b}.
Keep in mind that the entropy per particle $\sigma = S/N$ is assumed to be constant \cite{Lima_1992b,Lima_2014b}, i.e., $\dot{\sigma} =0$, where $ S \propto s a^3 $ is the entropy in the comoving volume. 
The constant value of $\sigma$ indicates $\dot{S} / S = \dot{N} / N = \Gamma$, which has been used for calculating the right-hand side of Eq.\ (\ref{eq:NonEquil_2}).
(Units are chosen so that $c = k_{B} =1$, unless otherwise stated. Here $k_{B}$ is the Boltzmann constant.)

The local Gibbs relation should be valid even for the nonequilibrium process considered here \cite{Lima_2014b}.
Accordingly, the thermodynamic quantities are related to the temperature $T$ by
\begin{equation}
         n k_{B} T d \left ( \frac{s}{n}    \right ) \equiv  n k_{B} T d\sigma  = d \varepsilon - \frac{\varepsilon + p}{n} dn      .
\label{eq:localGibbs}
\end{equation}
Substituting $\dot{\sigma} =0$ into Eq.\ (\ref{eq:localGibbs}), using both this result and Eq.\ (\ref{eq:NonEquil_1}), and rearranging Eq.\ (\ref{eq:NonEquil_3}), one obtains the dynamic creation pressure 
\begin{equation}
        p_{c}  = - (\varepsilon + p) \frac{\Gamma}{3 H}  .    
\label{eq:pc}
\end{equation}
The thermal evolution of matter creation can then be written as
\begin{equation}
       \frac{\dot{T}}{T}  =  \left ( \frac{\partial p}{\partial \varepsilon} \right )_{n}  \frac{\dot{n}}{n} =  \left ( \frac{\partial p}{\partial \varepsilon} \right )_{n} (\Gamma -3H) .
\label{eq:dotT_T}
\end{equation}
This equation has been derived by Lima \textit{et al.} \cite{Lima_2000,Lima_1992e,Lima_1992b,Lima_2014b}. 
(More general formulations have been examined in Ref.\ \cite{Lima_2000}. 
Harko has recently developed an equivalent formulation for a modified gravity theory with geometry-matter coupling \cite{Harko_2014}.)

For the nonequilibrium thermodynamic states considered here, a radiation temperature relation can be obtained from Eq.\ (\ref{eq:dotT_T}).
Substituting $p = \varepsilon / 3$ into Eq.\ (\ref{eq:dotT_T}), one finds
\begin{equation}
       \frac{\dot{T}}{T}  =    - \frac{\dot{a}}{a} +  \frac{\Gamma }{3}   \quad \textrm{(for radiation)} .
\label{eq:dotT_T_radiation}
\end{equation}
That can be rearranged as 
\begin{equation}
       \frac{1}{aT} \frac{d (aT)}{dt}  =  \frac{\Gamma }{3}  \quad  \textrm{or} \quad      \frac{d (aT)}{aT}  =  \frac{\Gamma }{3} dt .
\label{eq:dotT_T_radiation2}
\end{equation}
Integrating Eq.\ (\ref{eq:dotT_T_radiation2}) from arbitrary time $t$ to the present time $t_{0}$ gives
\begin{equation}
     \int_{aT}^{a_{0}T_{0}}  \frac{d (a^{\prime} T^{\prime} )}{a^{\prime} T^{\prime} }  =  \int_{t}^{t_{0}}   \frac{\Gamma (t^{\prime}) }{3} dt^{\prime}  , 
\label{eq:dotT_T_radiation_int1}
\end{equation}
and therefore
\begin{equation}
     \ln  \left ( \frac{ a_{0}T_{0} }{ aT } \right )   =  \frac{1}{3} \int_{t}^{t_{0}}   \Gamma (t^{\prime})  dt^{\prime}   
\label{eq:dotT_T_radiation_int2}
\end{equation}
where $a_{0}$ and $T_{0}$ are the present values of the scale factor and of the radiation temperature, respectively.
This equation can be rearranged as 
\begin{equation}
     T = T_{0}  \left ( \frac{ a_{0} }{ a } \right ) \exp  \left [   - \frac{1}{3} \int_{t}^{t_{0}}   \Gamma (t^{\prime})  dt^{\prime} \right ]  ,
\label{eq:dotT_T_radiation_T-t}
\end{equation}
so that
\begin{equation}
     T = T_{0}  \left ( \frac{ a_{0} }{ a } \right ) \exp  \left [   - \frac{1}{3} \int_{\tilde{a}}^{1}   \Gamma (\tilde{a}^{\prime})  \frac{dt^{\prime}}{d\tilde{a}^{\prime}} d\tilde{a}^{\prime} \right ]  ,
\label{eq:dotT_T_radiation_T-a}
\end{equation}
or equivalently
\begin{equation}
     T = T_{0}   ( 1 + z ) \exp  \left [    \frac{1}{3} \int_{0}^{z}   \Gamma (z^{\prime})  \frac{dt^{\prime}}{dz^{\prime}} dz^{\prime} \right ]   ,
\label{eq:dotT_T_radiation_T-z}
\end{equation}
where the normalized scale factor $\tilde{a}$ and the redshift $z$ are 
\begin{equation}
   \tilde{a} = \frac{a} { a_{0}}      \quad \textrm{and} \quad   z = \frac{ a_0 }{ a } -1   .
\label{eq:a_a0}
\end{equation}
Equations (\ref{eq:dotT_T_radiation_T-t}), (\ref{eq:dotT_T_radiation_T-a}), and (\ref{eq:dotT_T_radiation_T-z}) are the general radiative temperature law for adiabatic particle creation \cite{Lima_2014b}.
In this paper, that law is used to calculate the radiation temperature $T$ in a dissipative universe. 
When $\Gamma =0$, Eq.\ (\ref{eq:dotT_T_radiation_T-z}) reduces to the linear law $T = T_{0}(1+z)$.
However, $\Gamma$ is not zero for the nonequilibrium states considered here. 
Therefore, it is necessary to compute 
$\int_{\tilde{a}}^{1}   \Gamma (\tilde{a}^{\prime})  ( dt^{\prime} / d\tilde{a}^{\prime} ) d\tilde{a}^{\prime} $ in Eq.\ (\ref{eq:dotT_T_radiation_T-a})
and $\int_{0}^{z}   \Gamma (z^{\prime})  (dt^{\prime} / dz^{\prime} ) dz^{\prime} $ in Eq.\ (\ref{eq:dotT_T_radiation_T-z}) to calculate $T$.
These integrals depend on the limits, the particle production rate, and the background evolution of the universe.
In Sec.\ \ref{Results}, the calculated $T$ is compared with observations in the late universe. 
Accordingly, in the present study, the limits of integration in Eqs. (\ref{eq:dotT_T_radiation_T-t})--(\ref{eq:dotT_T_radiation_T-z}) correspond to the late universe.

A simple temperature relation is obtained from Eq.\ (\ref{eq:dotT_T_radiation}) for specific cases \cite{Lima_2000,Lima_2014b}.
For example, if $\Gamma = 3 \beta H = 3 \beta (\dot{a}/a)$, integrating Eq.\ (\ref{eq:dotT_T_radiation}) gives
\begin{equation}
       T = T_{0} \left ( \frac{a_{0}}{a} \right )^{1-\beta}   = T_{0}(1+z)^{1-\beta}         
\label{eq:dotT_T_radiation_T-z_beta}
\end{equation}
where $\beta$ is taken to be constant.
This simple temperature relation has been studied in detail \cite{Lima_2000}. 
In contrast, here a general radiative temperature law is considered for a dissipative universe.

\section{Modified dissipative model with constant terms}
\label{Dissipation model}

Entropic cosmology has been proposed to explain the accelerated expansion of the universe \cite{Easson12}.
In the entropic-force model, the horizon of the universe is assumed to have an associated entropy and an approximate temperature due to information holographically stored there \cite{Easson12}. 
Recently, various entropic-force models have been examined in detail \cite{Easson12,Cai1-Costa1,Basilakos1,Sola_2014a,Koma4,Koma5,Koma6,Koma7}.
For example, in place of the Bekenstein entropy \cite{Bekenstein1}, the Tsallis-Cirto entropy \cite{Tsallis2012} based on nonextensive statistics \cite{Tsa0} has been applied to the horizon of the universe \cite{Koma5,Koma6}. 
Basilakos \textit{et al.} \cite{Basilakos1,Sola_2014a} have shown that simple combinations of pure Hubble terms, such as $H^{2}$, $\dot{H}$, and $H$, are insufficient for a complete description of the cosmological data \cite{Koma7}. 
Therefore, the constant term plays an important role.
Accordingly, a phenomenological entropic-force model that includes constant terms, an irreversible entropy $S_{\textrm{irr}}$, and a kind of reversible entropy $S_{\textrm{rev}}$ has been proposed \cite{Koma7}.
Using this model, a dissipative universe can be analyzed systematically, over the entire range from a nondissipative $\Lambda$CDM model to a fully dissipative CCDM model.
In the present work, it is called the modified dissipative model.
In what follows, it is briefly reviewed according to Ref.\ \cite{Koma7}.
Keep in mind that the entropic-force considered here is different from the idea that gravity itself is an entropic force \cite{Padma1,Verlinde1}.

The Friedmann, acceleration, and continuity equations for the modified dissipative model become
\begin{equation}
   H^{2}  = \frac{ 8\pi G }{ 3 } \rho     +  \alpha  H_{0}^{2}   ,
\label{eq:FRW01(g)}
\end{equation}
\begin{align} 
  \frac{ \ddot{a}  }{ a }    &=  -  \frac{ 4\pi G }{ 3 } \left ( \rho  + \frac{3p}{c^2}  \right )                  +  \alpha  H_{0}^{2} +  \gamma_{\textrm{irr}}  H_{0}^{2}      \notag \\
                                    &=  -  \frac{ 4\pi G }{ 3 } \left ( \rho  + \frac{3           p_{e}  }{c^2}  \right )  + \alpha  H_{0}^{2}                        ,   
\label{eq:FRW02(g)}
\end{align} 
and
\begin{equation}
       \dot{\rho} + 3  \frac{\dot{a}}{a} \left (  \rho + \frac{p_{e}}{c^2}  \right )   = 0   ,
\label{eq:drho_General_01_pe}
\end{equation}
with two dimensionless constants
\begin{equation}
   \alpha     \geq 0   \quad  \textrm{and}  \quad     \gamma_{\textrm{irr}}    \geq 0   .
\end{equation}
Here $G$, $H_{0}$, and $\rho$ are the gravitational constant, the Hubble parameter at the present time, and the \textit{mass} density of the cosmological fluid, respectively \cite{Koma7}. 
The mass density is $\rho \equiv \varepsilon /c^{2}$.
The effective pressure $p_{e}$ in Eqs. (\ref{eq:FRW02(g)}) and (\ref{eq:drho_General_01_pe}) is 
\begin{equation}
    p_{e}   =  p + p_{c}  
\end{equation}
where $p_{c}$ is a pressure derived from irreversible entropy related to dissipative processes.
In this study, $p_{c}$ is taken to be equivalent to the dynamic creation pressure in the CCDM model.
(For $\Lambda (t)$CDM models, nonzero terms related to reversible entropy appear on the right-hand side of the continuity equation \cite{Koma7}.
That model involves the transfer of energy between two fluids \cite{Barrow22,Amendola1Zimdahl01,Wang0102_YWang2014,Pavon_2005}.)

The modified dissipative model is used to calculate a radiation temperature $T$ from the general radiative temperature law in Eqs. (\ref{eq:dotT_T_radiation_T-t})--(\ref{eq:dotT_T_radiation_T-z}).
In Sec.\ \ref{Results}, the $T$--$z$ relation in the late universe is examined, to compare the calculated value of $T$ with observations.
The pressure of the cosmological fluid in the present model is negligible, $p = 0$.
A matter-dominated universe is assumed.
Consequently, the effective pressure $p_{e}$ is 
\begin{equation}
   p_{e} =  p + p_{c} = p_{c} = - \frac{  c^{2} H_{0}^{2}} {  4\pi G }  \gamma_{\textrm{irr}}  .
\label{eq:pe_Q_1_alpha4}
\end{equation}
On the other hand, substituting $p_{c} = p_{e}$, $p = 0$, and $\varepsilon = \rho c^{2}$ into Eq.\ (\ref{eq:pc}), one obtains
\begin{equation} 
  p_{e}   =   - \rho c^{2}   \frac{\Gamma}{3H}       . 
\label{eq:pe_Gamma}
\end{equation}
Accordingly, from Eqs.\ (\ref{eq:pe_Q_1_alpha4}) and (\ref{eq:pe_Gamma}), 
\begin{equation} 
    \Gamma = \frac{3H}{4 \pi G}  \frac{  \gamma_{\textrm{irr}}  H_{0}^{2} }{ \rho }        .
\label{eq:Gamma_0_alpha4}
\end{equation}

Assume that the $\alpha  H_{0}^{2}$ terms in Eqs.\ (\ref{eq:FRW01(g)}) and (\ref{eq:FRW02(g)}) are equivalent to the cosmological constant $\Lambda /3$ in the standard $\Lambda$CDM model.
That is, $dS_{\textrm{rev}}=0$ for homogeneous systems in CCDM models \cite{Prigogine_1988b,Harko_2014}. 
In contrast, $p_{e}$ is related to the irreversible entropy in dissipative processes because $p_{e} =p_{c}$.
In other words, the $\gamma_{\textrm{irr}}  H_{0}^{2}$ term in Eq.\ (\ref{eq:FRW02(g)}) is related to the irreversible entropy.
Accordingly, the preceding cosmological equations are equivalent to those for an extended $\Lambda$CDM model in a dissipative universe.
Therefore, the $\alpha  H_{0}^{2}$ term is interpreted as a modification of the Einstein tensor.
In contrast, $p_{e}$ is a modification of the energy--momentum tensor of the Einstein equation  \cite{Koma7}.
The properties of this model are equivalent to those of the modified entropic-force model examined in Ref.\ \cite{Koma7}.

Combining Eqs.\ (\ref{eq:FRW01(g)}) and (\ref{eq:FRW02(g)}), and using Eq.\ (\ref{eq:pe_Q_1_alpha4}), one finds  
\begin{equation}
 \dot{H}    = -  C_{m} H^2    +   C_{ag}  H_{0}^{2}     
\label{eq:dHC1C3hC4h(Cst)}
\end{equation}
where $C_{m}$ and $C_{ag}$ are dimensionless constants \cite{Koma7} 
\begin{equation}
     C_{m} = 1.5   \quad   \textrm{and}  \quad  {C}_{ag}  =  \frac{ 3 \alpha + 2 \gamma_{\textrm{irr}}  }{   2   }    .
\label{eq:Cm=1.5_C4}
\end{equation}
Here $C_{m} =1.5$ corresponds to a matter-dominated universe in standard cosmology \cite{Weinberg1,Roy1}.
Solving Eq.\ (\ref{eq:dHC1C3hC4h(Cst)}), one obtains 
\begin{align}
 \left ( \frac{H} {H_{0}} \right )^{2}   &=  \left ( 1-  \tilde{\Omega}_{\Lambda} \right )   \tilde{a}^{ - 3}   +  \tilde{\Omega}_{\Lambda}   \notag \\
                                                    &=                                      \tilde{\Omega}_{m}    \tilde{a}^{ - 3}   +  \tilde{\Omega}_{\Lambda}   
\label{eq:H/H0(C1C4)_02}
\end{align}
where $\tilde{a}$ = $ a / a_{0}$ in Eq.\ (\ref{eq:a_a0}).
The two constant parameters are defined as 
\begin{equation}
 \tilde{\Omega}_{\Lambda} \equiv  \frac{ C_{ag} }{ C_{m}  }   \quad   \textrm{and}   \quad  \tilde{\Omega}_{m}  \equiv  1 - \tilde{\Omega}_{\Lambda} .
\label{eq:tOmega_L}
\end{equation}
This solution is the same as that in the standard $\Lambda$CDM model \cite{Koma6,Koma7}. 
Accordingly, the constant term $\tilde{\Omega}_{\Lambda}$ behaves as if it were $\Omega_{\Lambda}$ in the standard $\Lambda$CDM model.
Similarly, $\tilde{\Omega}_{m}$ behaves as if it were $\Omega_{m}$.
Here $\Omega_{m}$ and $\Omega_{\Lambda}$ are the density parameters for matter and for $\Lambda$, respectively. 
The density parameter $\Omega_{r}$ for radiation is neglected in the late universe.

To study a dissipative universe quantitatively, define a dissipation rate \cite{Koma7}  
\begin{equation}
      \tilde{\mu}  \equiv  \frac{ \gamma_{\textrm{irr}} }{ C_{ag}  }     =  \frac{ \tilde{\Omega}_{D} }{ \tilde{\Omega}_{\Lambda} }      
\label{eq:dissipation_0}
\end{equation}
where $\tilde{\Omega}_{D}$ is a constant parameter related to dissipative processes,
\begin{equation}
      \tilde{\Omega}_{D}  \equiv  \frac{ \gamma_{\textrm{irr}} }{ C_{m} }      .
\label{eq:Omega_D_0}
\end{equation}
As discussed in Ref.\ \cite{Koma7}, when $\gamma_{\textrm{irr}} = 0 $, one obtains $\tilde{\mu} =0$ from Eq.\ (\ref{eq:dissipation_0}).
In this case, the present model is equivalent to the standard nondissipative $\Lambda$CDM model.
In contrast, when $\alpha =0$, one obtains $\tilde{\mu} =1$ from Eq.\ (\ref{eq:dissipation_0}) because $C_{ag} = \frac{ 3 \alpha + 2 \gamma_{\textrm{irr}}  }{2} = \gamma_{\textrm{irr}} $.
In that case, the present model is equivalent to the fully dissipative CCDM model proposed by Lima \textit{et al.} \cite{Lima2010}.
That is, $\tilde{\mu} =0$ corresponds to the nondissipative $\Lambda$CDM model, whereas $\tilde{\mu} =1$ corresponds to the fully dissipative CCDM model.
In this way, the extent of the dissipative universe is determined by the dissipation rate $\tilde{\mu}$.

Next, consider the effective equation-of-state parameter in the modified dissipative model. 
From Eq.\ (\ref{eq:pe_Gamma}), $w_{e}$ becomes
\begin{equation}
  w_{e} \equiv \frac{p_{e}}{\rho c^{2}}      = - \frac{\Gamma}{3H}  .
\label{eq:we_p_alpha4}
\end{equation}
After some algebra, one finds 
\begin{equation}
w_{e} = - \frac{  \tilde{\Omega}_{D}  \tilde{a}^{3}    }{    1-  \tilde{\Omega}_{\Lambda}     +  \tilde{\Omega}_{D}  \tilde{a}^{3}     } 
        = - \frac{  \tilde{\Omega}_{D}  \tilde{a}^{3}    }{                   \tilde{\Omega}_{m}     +  \tilde{\Omega}_{D}  \tilde{a}^{3}     }   
\label{eq:we-a_alpha4}
\end{equation}
where $\tilde{a}$ is the normalized scale factor $a/a_{0}$. 
For details, see Ref.\ \cite{Koma7}.
Note that $w_{e}$ is not equal to the equation-of-state parameter $w$ for a generic component of matter, because $w$ is zero in a matter-dominated universe, for which $p=0$.
(A more general inhomogeneous equation of state has been examined in Ref.\ \cite{Odintsov_2006_b}.)

\section{Radiation temperature in the modified dissipative model}
\label{Temperature law for the dissipation model}

The general radiative temperature law in Eq.\ (\ref{eq:dotT_T_radiation_T-a}) is rearranged as
\begin{align}
     T  &= T_{0}  \left ( \frac{ a_{0} }{ a } \right ) \exp  \left [   - \frac{1}{3} \int_{\tilde{a}}^{1}   \Gamma (\tilde{a}^{\prime})  \frac{dt^{\prime}}{d\tilde{a}^{\prime}} d\tilde{a}^{\prime} \right ]   \notag \\
        &= T_{0}  ( 1 + z )                                   \exp  \left [   - \frac{1}{3} \int_{\tilde{a}}^{1}   \Gamma (\tilde{a}^{\prime})  \frac{dt^{\prime}}{d\tilde{a}^{\prime}} d\tilde{a}^{\prime} \right ]                  .
\label{eq:dotT_T_radiation_T-a_z}
\end{align}

First consider the calculation of $dt/d\tilde{a}$ in Eq.\ (\ref{eq:dotT_T_radiation_T-a_z}). 
(For simplicity, the prime is omitted.)
As discussed in Sec.\ \ref{Dissipation model}, the background evolution in the modified dissipative model is the same as that in the standard $\Lambda$CDM model. 
Accordingly, solutions of the standard $\Lambda$CDM model are used for a spatially flat universe.
The solution \cite{Sola_2009,Matsubara1} is
\begin{equation}
 \tilde{a} (t) =  \left (   \frac{ \tilde{\Omega}_{m}  }{  \tilde{\Omega}_{\Lambda}  } \right )^{1/3}   \sinh^{2/3} \left (    \frac{  3 H_{0} \sqrt{ \tilde{\Omega}_{\Lambda} } t }{2}    \right )    
\label{eq:a-t_1}
\end{equation}
and equivalently
\begin{equation}
  t (\tilde{a}) =  \frac{2}{  3 H_{0} \sqrt{ \tilde{\Omega}_{\Lambda} }  }    \sinh^{-1} \left (    \sqrt { \frac{ \tilde{\Omega}_{\Lambda}  }{  \tilde{\Omega}_{m}  } }  \tilde{a}^{3/2} \right )   .
\label{eq:t-a_1}
\end{equation}
The density parameter for radiation is negligible in this late universe.
Differentiating Eq.\ (\ref{eq:a-t_1}) with respect to $t$, substituting Eq.\ (\ref{eq:t-a_1}) into the result, and rearranging, one obtains 
\begin{equation}
   \frac{dt}{ d \tilde{a} } = \frac{  \sinh^{1/3} g(\tilde{a})  \cosh^{-1} g(\tilde{a})  }{  H_{0} \tilde{\Omega}_{\Lambda}^{1/6}  \tilde{\Omega}_{m}^{1/3}    }   
\label{eq:dtda_1}
\end{equation}
where 
\begin{equation}
    g(\tilde{a}) = \sinh^{-1}  \left (  \sqrt{   \frac{\tilde{\Omega}_{\Lambda}}{\tilde{\Omega}_{m}}  }   \tilde{a}^{3/2}    \right )    .
\label{eq:g(a)_1}
\end{equation}
Next, rearrange $\Gamma$ in Eq.\ (\ref{eq:dotT_T_radiation_T-a_z}).
Using Eq.\ (\ref{eq:we_p_alpha4}), it can be written as
\begin{equation}
\Gamma = - 3H w_{e} = - 3 \left ( \frac{H}{H_{0}} \right ) H_{0} w_{e}    .
\label{eq:Gamma3Hwe_1}
\end{equation}
Substituting Eqs.\ (\ref{eq:H/H0(C1C4)_02}) and (\ref{eq:we-a_alpha4}) into Eq.\ (\ref{eq:Gamma3Hwe_1}) leads to 
\begin{equation}
 \Gamma =  \frac{   3  H_{0} \tilde{\Omega}_{D}  \sqrt{ \tilde{\Omega}_{m}  \tilde{a}^{ - 3}   +  \tilde{\Omega}_{\Lambda} }   }{  \tilde{\Omega}_{m}  \tilde{a}^{-3} +  \tilde{\Omega}_{D}   }   .
\label{eq:Gamma_2}
\end{equation}

Substituting Eqs.\ (\ref{eq:dtda_1}) and (\ref{eq:Gamma_2}) into Eq.\ (\ref{eq:dotT_T_radiation_T-a_z}), one finds  
\begin{align}
T  = & T_{0}  (  1 + z )           \exp     \left [   - \frac{1}{3} \int_{\tilde{a}}^{1}   \Gamma (\tilde{a}^{\prime})  \frac{dt^{\prime}}{d\tilde{a}^{\prime}} d\tilde{a}^{\prime} \right ]   \notag \\
    = & T_{0}  (  1 + z )
                                          \exp     \Biggl[  \frac{  - \tilde{\Omega}_{D}   }{  \tilde{\Omega}_{\Lambda}^{1/6}   \tilde{\Omega}_{m}^{1/3}   }    \int_{\tilde{a}}^{1}  K(\tilde{a}^{\prime})   d\tilde{a}^{\prime}  \Biggr]                       
\label{eq:dotT_T_radiation_T-a_z_2}
\end{align}
where 
\begin{equation}
  K(\tilde{a})  =  \frac{   \sqrt{ \tilde{\Omega}_{m}  \tilde{a}^{ - 3}   +  \tilde{\Omega}_{\Lambda} }   }{  \tilde{\Omega}_{m}  \tilde{a}^{-3} +  \tilde{\Omega}_{D}   }     \sinh^{1/3} g(\tilde{a})  \cosh^{-1} g(\tilde{a})       
\label{eq:K}
\end{equation}
and $g(\tilde{a})$ is given by Eq.\ (\ref{eq:g(a)_1}).
Equation\ (\ref{eq:dotT_T_radiation_T-a_z_2}) is the radiation temperature--redshift relation for the modified dissipative model in the late universe.
The influence of dissipation is included in $\tilde{\Omega}_{D}$.
Using Eq.\ (\ref{eq:dotT_T_radiation_T-a_z_2}), the radiation temperature can be determined as a function of the dissipation rate  $\tilde{\mu}$ 
where $\tilde{\mu}$ = $\tilde{\Omega}_{D} / \tilde{\Omega}_{\Lambda}$  from Eq.\ (\ref{eq:dissipation_0}).

As a specific case, a simple temperature relation can be obtained from Eq.\ (\ref{eq:dotT_T_radiation}). 
For example, if $w_{e} = - \Gamma / (3H)$ from Eq.\ (\ref{eq:we_p_alpha4}) is assumed to be constant, then 
\begin{equation}
       T = T_{0}(1+z)^{1 + w_{e}}           .
\label{eq:dotT_T_radiation_T-z_we}
\end{equation}
This equation is equivalent to Eq.\ (\ref{eq:dotT_T_radiation_T-z_beta}), replacing $w_{e}$ by $ - \beta$.
In general, $w_{e}$ is not constant in a dissipative universe, as examined in the next section.
Accordingly, Eq.\ (\ref{eq:dotT_T_radiation_T-a_z_2}) plays an important role in studying the dissipative universe.

\section{Evolution of the radiation temperature in a dissipative universe}
\label{Results}

\begin{figure} [b] 
\begin{minipage}{0.495\textwidth}
\begin{center}
\scalebox{0.31}{\includegraphics{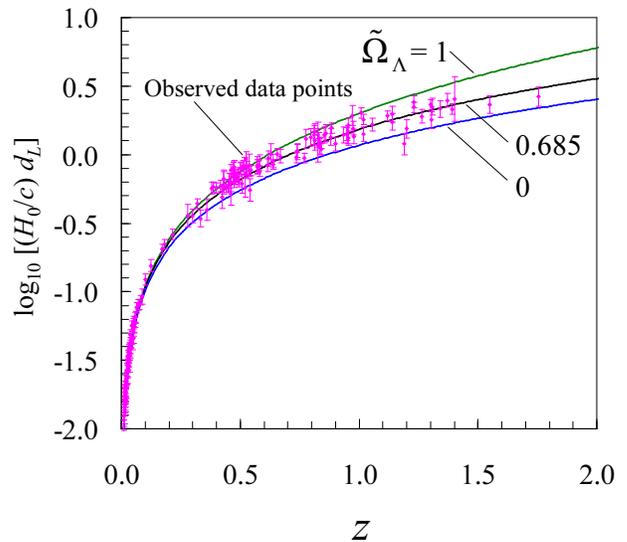}}
\end{center}
\end{minipage}
\caption{ (Color online). Dependence of the luminosity distance $d_L$ on the redshift $z$. 
The three continuous curves represent the modified dissipative model for $\tilde{\Omega}_{\Lambda} = 0$, $0.685$, and $1$.
They are respectively equivalent to the standard $\Lambda$CDM model for $(\Omega_{m}, \Omega_{\Lambda}) = (1, 0), (0.315, 0.685)$, and $(0, 1)$ in a spatially flat universe.
The closed diamonds with error bars are supernova data \cite{Riess2007SN1}, for which $H_{0}$ is $67.3$ km/s/Mpc based on Planck 2013 results \cite{Planck2013}. 
A similar $d_L$--$z$ relation has been discussed in entropic cosmology \cite{Koma4,Koma5,Koma6}.   }
\label{Fig-dL-z}
\end{figure}

\begin{figure} [t] 
\begin{minipage}{0.495\textwidth}
\begin{center}
\scalebox{0.3}{\includegraphics{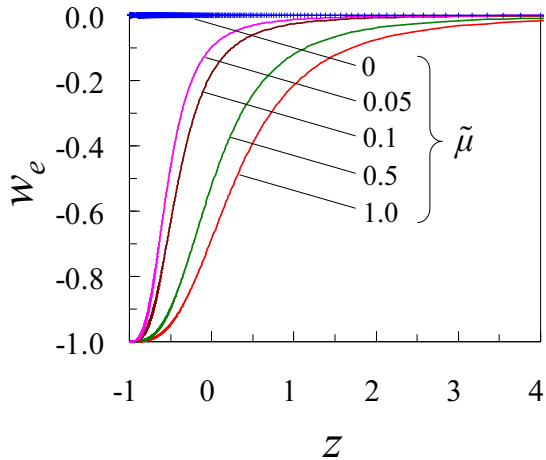}}
\end{center}
\end{minipage}
\caption{ (Color online). Dependence of the effective equation-of-state parameter $w_{e}$ on the redshift $z$ for the indicated dissipation rates $\tilde{\mu}$. 
Zero dissipation corresponds to a nondissipative $\Lambda$CDM model, whereas $\tilde{\mu} =1$ corresponds to a fully dissipative CCDM model.
The background evolution of the universe in each case is equivalent to that in the fine-tuned standard $\Lambda$CDM model because $\tilde{\Omega}_{\Lambda}  = \Omega_{\Lambda} = 0.685$.   
The dependence of $w_e$ on $\tilde{a}$ has been discussed in Ref.\ \cite{Koma7}.     } 
\label{Fig-we-z}
\end{figure}

\begin{figure} [t] 
\begin{minipage}{0.495\textwidth}
\begin{center}
\scalebox{0.3}{\includegraphics{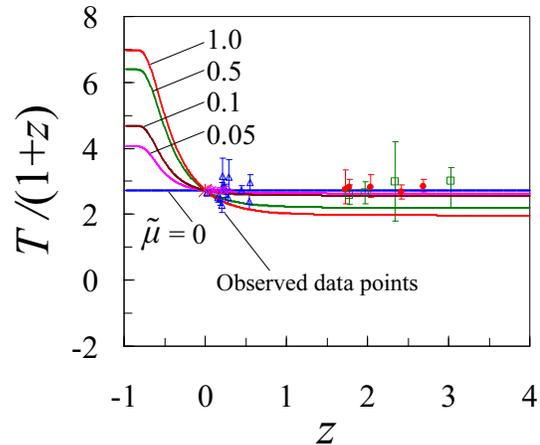}}
\end{center}
\end{minipage}
\caption{ (Color online). The radiation temperature--redshift relation.
The vertical axis is $T/ (1+z)$.
The continuous curves represent the modified dissipative model for $\tilde{\mu} =0$, $0.05$, $0.1$, $0.5$, and $1$. 
The background evolution of the universe for each value of $\tilde{\mu}$ is the same because $\tilde{\Omega}_{\Lambda}  = 0.685$.  
The symbols with error bars are observed data points \ \cite{Mather_1999,Ge_1997,Srianand_2000,Molaro_2002,Cui_2005,Srianand_2008,Luzzi_2009,Noterdaeme_2010,Noterdaeme_2011,Martino_2015}.
The original data represented by the green squares, blue triangles, and red circles are from Refs.\ \cite{Ge_1997,Srianand_2000,Molaro_2002,Cui_2005}, \cite{Luzzi_2009}, and \cite{Srianand_2008,Noterdaeme_2010,Noterdaeme_2011}, respectively.
The recent data represented by the pink diamonds at low redshifts are from Ref.\ \cite{Martino_2015}.
The data point for $z=0$ is from Ref. \cite{Mather_1999}.   
For the observed data points, see Table \ref{tab-T-z_obs} and Ref.\ \cite{Noterdaeme_2011}. 
Note that several error bars shown here are asymmetric because the corresponding errors shown in Table \ref{tab-T-z_obs} have been asymmetric.  }
\label{Fig-T-z}
\end{figure}

\begin{table}[t]
\caption{ The observed temperature--redshift relation. 
The measured values are based on the analysis of a fine structure of atomic carbon \cite{Ge_1997,Srianand_2000,Molaro_2002,Cui_2005}, the Sunyaev-Zel'dovich effect \cite{Luzzi_2009}, the rotational excitation of CO molecules \cite{Srianand_2008,Noterdaeme_2010,Noterdaeme_2011}, 
and a sample of X-ray selected clusters \cite{Martino_2015}.
The temperature at $z = 0$ is taken from Ref.\ \cite{Mather_1999}.     }
\label{tab-T-z_obs}
\newcommand{\m}{\hphantom{$-$}}
\newcommand{\cc}[1]{\multicolumn{1}{c}{#1}}
\renewcommand{\tabcolsep}{1.5pc} 
\renewcommand{\arraystretch}{1.15} 
\begin{tabular}{@{}lllll}
\hline
  $z$        &  $T \quad (\textrm{K})$                  &  $\textrm{Refs.}$      \\
\hline
\hline
$0.000$    & $2.725 \pm 0.002$          &  \cite{Mather_1999} \\
\hline
$0.023$    & $2.72   \pm 0.10$            &  \cite{Luzzi_2009} \\
$0.152$    & $2.90   \pm 0.17$            &  \cite{Luzzi_2009} \\
$0.183$    & $2.95   \pm 0.27$            &  \cite{Luzzi_2009} \\
$0.200$    & $2.74   \pm 0.28$            &  \cite{Luzzi_2009} \\
$0.202$    & $3.36   \pm 0.20$            &  \cite{Luzzi_2009} \\
$0.216$    & $3.85   \pm 0.64$            &  \cite{Luzzi_2009} \\
$0.232$    & $3.51   \pm 0.25$            &  \cite{Luzzi_2009} \\
$0.252$    & $3.39   \pm 0.26$            &  \cite{Luzzi_2009} \\
$0.282$    & $3.22   \pm 0.26$            &  \cite{Luzzi_2009} \\
$0.291$    & $4.05   \pm 0.66$            &  \cite{Luzzi_2009} \\
$0.451$    & $3.97   \pm 0.19$            &  \cite{Luzzi_2009} \\
$0.546$    & $3.69   \pm 0.37$            &  \cite{Luzzi_2009} \\
$0.550$    & $4.59   \pm 0.36$            &  \cite{Luzzi_2009} \\
\hline
$0.042$     & $2.857   \pm 0.018$        &  \cite{Martino_2015} \\
$0.077$     & $2.953   \pm 0.021$        &  \cite{Martino_2015} \\
$0.123$     & $3.072   \pm 0.026$        &  \cite{Martino_2015} \\
$0.169$     & $3.162   \pm 0.020$        &  \cite{Martino_2015} \\
$0.222$     & $3.326   \pm 0.015$        &  \cite{Martino_2015} \\
$0.274$     & $3.495   \pm 0.016$        &  \cite{Martino_2015} \\
\hline
$1.729$    & $7.5^{+ 1.6}_{-1.2}$           &  \cite{Noterdaeme_2011} \\
$1.774$    & $7.8^{+0.7}_{-0.6}$            &  \cite{Noterdaeme_2011} \\
$2.038$    & $8.6^{+1.1}_{-1.0}$            &  \cite{Noterdaeme_2011} \\
$2.418$    & $9.15   \pm   0.7$            &  \cite{Srianand_2008,Noterdaeme_2011} \\
$2.690$    & $10.5^{+0.8}_{-0.6}$          &  \cite{Noterdaeme_2010,Noterdaeme_2011} \\
\hline
$1.777$      & $7.2   \pm  0.8$            &   \cite{Cui_2005}         \\
$1.973$      & $7.9   \pm  1.0$            &   \cite{Ge_1997}          \\
$2.337$      & $10    \pm  4$              &   \cite{Srianand_2000} \\
$3.025$      & $12.1^{+1.7}_{-3.2}$       &   \cite{Molaro_2002}    \\
\hline
\hline
\end{tabular}\\
 \end{table}

The dissipation rate $\tilde{\mu}$ from Eq.\ (\ref{eq:dissipation_0}) is 
\begin{equation}
      \tilde{\mu}  \equiv  \frac{ \gamma_{\textrm{irr}} }{ C_{ag}  }     =  \frac{ \tilde{\Omega}_{D} }{ \tilde{\Omega}_{\Lambda} }      .
\label{eq:dissipation_1}
\end{equation}
Zero dissipation corresponds to a nondissipative $\Lambda$CDM model, whereas $\tilde{\mu} =1$ corresponds to a fully dissipative CCDM model.
As discussed in Ref.\ \cite{Koma7}, $\tilde{\Omega}_{\Lambda}$ can be determined from the background evolution of the universe.
Accordingly, $\tilde{\Omega}_{\Lambda}$ = $\Omega_{\Lambda}$ from a fine-tuned standard $\Lambda$CDM model.
Consider a spatially flat universe in which $(\Omega_{m}, \Omega_{\Lambda}) = (0.315, 0.685)$ based on the Planck 2013 results \cite{Planck2013}.
That is, let $\tilde{\Omega}_{\Lambda} =\Omega_{\Lambda}  = 0.685$ \cite{Koma7}.
To confirm the background evolution of the universe in the present dissipative model, consider the luminosity distance $d_{L}$ \cite{Sato1} given by 
\begin{equation}
  \left ( \frac{ H_{0} }{ c } \right )   d_{L}      =   (1+z)  \int_{1}^{1+z}  \frac{dy} { F(y) }    .
\label{eq:dL_00}  
\end{equation}
The integrating variable $y$ and the function $F(y)$ are given by 
\begin{equation}
  y = \frac{a_0} {a} = \tilde{a}^{-1}  \quad \textrm{and} \quad    F(y)  = \frac{ H }{ H_{0} } 
\end{equation}
where $H/H_{0}$ is from Eq.\ (\ref{eq:H/H0(C1C4)_02}).
As shown in Fig.\ \ref{Fig-dL-z}, it is found that $d_{L}$ for $\tilde{\Omega}_{\Lambda} = 0.685$ agrees with the supernova data.
($\tilde{\Omega}_{\Lambda}  = 0.685$ used here is approximately equivalent to $\Omega_{\Lambda}  = 0.691$ for the recent Planck 2015 results \cite{Planck2015}.
The influence of $\tilde{\Omega}_{\Lambda}$ is investigated later.)

To examine the influence of the dissipation rate, $\tilde{\mu}$ is set to several typical values, $0$, $0.05$, $0.1$, $0.5$, and $1.0$ in turn.
The background evolution of the universe in each case is equivalent to that in the fine-tuned standard $\Lambda$CDM model.
As shown in Fig.\ \ref{Fig-dL-z}, the background evolution agrees with the observed supernova data because $\tilde{\Omega}_{\Lambda} =\Omega_{\Lambda}  = 0.685$.

Before studying the radiation temperature, the evolution of the effective equation-of-state parameter $w_{e}$ is examined.
To this end, $w_e$ for various values of $\tilde{\mu}$ is plotted in Fig.\ \ref{Fig-we-z}, where $w_e$ is calculated from Eqs.\ (\ref{eq:we-a_alpha4}) and (\ref{eq:a_a0}). 
In this figure, $\tilde{\mu} =0$ corresponds to a nondissipative $\Lambda$CDM model, whereas $\tilde{\mu} =1$ corresponds to a fully dissipative CCDM model.
As shown in Fig.\ \ref{Fig-we-z}, $w_{e}$ for $\tilde{\mu} =0$ is always equal to $0$ because $p_{e} = 0$.
However, $w_{e}$ decreases with increasing $\tilde{\mu}$.
It is found that the dissipation rate $\tilde{\mu}$ affects $w_{e}$ even if the background evolution of the universe is not altered.
In addition, $w_{e}$ for $\tilde{\mu} >0$ gradually decreases with decreasing $z$ and eventually approaches $-1$.
That is, $w_{e}$ is not constant when $\tilde{\mu} >0$.
The varying value of $w_{e}$ indicates that the simple $T$--$z$ relation from Eq.\ (\ref{eq:dotT_T_radiation_T-z_we}) is not suited to describe the radiation temperature in the present model. 
Accordingly, Eq.\ (\ref{eq:dotT_T_radiation_T-a_z_2}) plays an important role.

Next, the radiation temperature in the modified dissipative model is examined for various values of $\tilde{\mu}$.
The radiation temperature $T$ in the late universe is numerically calculated from Eq.\ (\ref{eq:dotT_T_radiation_T-a_z_2}).
For this calculation, the redshift is varied from $z= -0.98$ to $4$, i.e., the normalized scale factor is varied between $\tilde{a} =50$ and $\tilde{a} =0.2$. 
To compare with observations, the data points are taken from Refs.\ \cite{Mather_1999,Ge_1997,Srianand_2000,Molaro_2002,Cui_2005,Srianand_2008,Luzzi_2009,Noterdaeme_2010,Noterdaeme_2011,Martino_2015}.  
(The values are summarized in Table \ref{tab-T-z_obs}. 
For details, see, e.g., Ref.\ \cite{Noterdaeme_2011}.)
In Fig.\ \ref{Fig-T-z}, the vertical axis is $T/ (1+z)$ so that a horizontal line can correspond to a linear law $T= T_{0} (1+z)$ in equilibrium.
From this figure, it is found that the radiation temperature $T$ for $\tilde{\mu} =0$ obeys the linear law and is consistent with the observed data.
However, with increasing dissipation $\tilde{\mu}$, $T$ gradually deviates from both the linear law and the observations.
In particular, $T$ for $\tilde{\mu} =1$ is far off the observed data points, due to the nonequilibrium dissipative processes.
Accordingly, a fully dissipative universe is constrained even if the background evolution of the universe is the same.
On the other hand, $T$ for $\tilde{\mu} =0.05$ agrees with the observed data points.
This agreement implies that a weakly dissipative universe fits the observed $T$-$z$ relation.
The weakly dissipative universe is discussed later.

In the preceding discussion, $\tilde{\Omega}_{\Lambda} = \Omega_{\Lambda} = 0.685$.
Accordingly, the background evolution of the universe is the same.
Consequently, a weakly dissipative model is consistent with observations.
Finally, to examine the influence of both $\tilde{\mu}$ and $\tilde{\Omega}_{\Lambda}$, a likelihood analysis for the radiation temperature is performed.
(A similar analysis was performed in Ref.\ \cite{Koma7}, in which a growth rate for clustering was examined.)
For this purpose, $\tilde{\mu}$ and $\tilde{\Omega}_{\Lambda}$ are treated as free parameters.
The chi-squared function for the radiation temperature then becomes 
\begin{equation}
\chi^{2}_{\textrm{RT}} (\tilde{\Omega}_{\Lambda}, \tilde{\mu})
= \sum\limits_{i=1}^{29} { \left[ \frac{   
                                                         T_{\textrm{obs}} (z_{i} ) - T_{\textrm{cal}} (z_{i}, \tilde{\Omega}_{\Lambda}, \tilde{\mu} )  }{ \sigma_{i}^{\textrm{RT}} }   
                                     \right]^{2}   }   
\label{eq:chi_mu}
\end{equation}
where $ T_{\textrm{obs}} (z_{i} )$ and $ T_{\textrm{cal}} (z_{i}, \tilde{\Omega}_{\Lambda}, \tilde{\mu} ) $ are the observed and calculated radiation temperatures, respectively, and $\sigma_{i}^{\textrm{RT}}$ is the uncertainty in the observed temperature.
The observed data points (numbered $i=1$ to $29$) are summarized in Table\ \ref{tab-T-z_obs}. 
For the likelihood analysis, $\tilde{\Omega}_{\Lambda}$ and $\tilde{\mu}$ are sampled in the range 0 to 1 in steps of $0.005$.
Therefore, negative dissipation rates are not considered. 
Using $\chi^{2}_{\textrm{RT}}$ from Eq.\ (\ref{eq:chi_mu}), the likelihood function $L_{\textrm{RT}}$ is \cite{Lima2010} 
\begin{equation}
 L_{\textrm{RT}} \propto \exp ({- \chi^{2}_{\textrm{RT}} / 2}) . 
\label{eq:L-chi}
\end{equation}
For simplicity, $L_{\textrm{RT}}$ is normalized.
Note that $\tilde{\Omega}_{\Lambda} =0$ and $\tilde{\Omega}_{\Lambda} = 1$ have not been sampled, in order to avoid a division by zero when $T$ is calculated from Eq.\ (\ref{eq:dotT_T_radiation_T-a_z_2}). 
Here $\tilde{\Omega}_{\Lambda} =1$ corresponds to $\tilde{\Omega}_{m} =0$ because $\tilde{\Omega}_{m} = 1 - \tilde{\Omega}_{\Lambda}$.

\begin{figure} [t] 
\begin{minipage}{0.495\textwidth}
\begin{center}
\scalebox{0.33}{\includegraphics{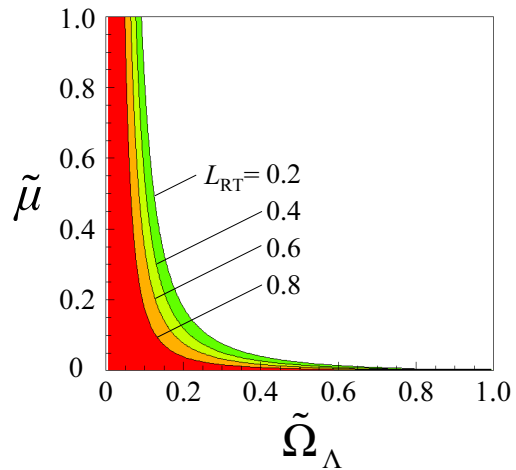}}
\end{center}
\end{minipage}
\caption{ (Color online). Contours of the normalized likelihood $L_{\textrm{RT}}$ in the $(\tilde{\Omega}_{\Lambda}, \tilde{\mu})$ plane for the radiation temperature.
The contours of $L_{\textrm{RT}}$ for $0.2$, $0.4$, $0.6$, and $0.8$ are plotted. 
The likelihood function is normalized using the maximum value, obtained for $(\tilde{\Omega}_{\Lambda}, \tilde{\mu}) = (\tilde{\Omega}_{\Lambda}, 0)$. 
Note that the maximum value is the same for all values of $\tilde\Omega_\Lambda$ when $\tilde\mu =0$.  }
\label{Fig--L_mu_Lambda-T}
\end{figure}

\begin{figure} [t] 
\begin{minipage}{0.495\textwidth}
\begin{center}
\scalebox{0.33}{\includegraphics{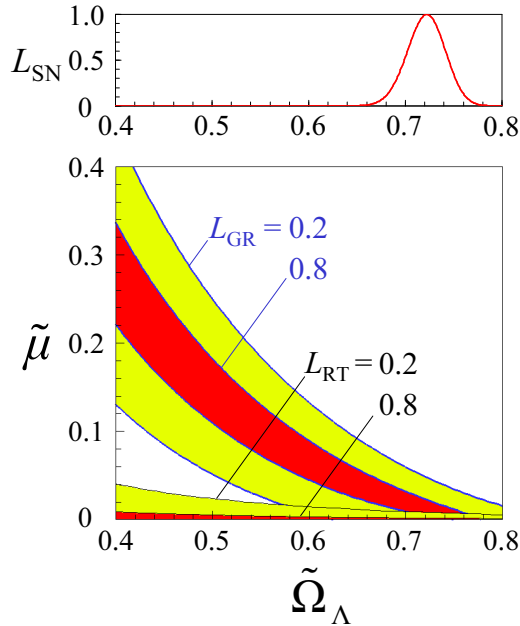}}
\end{center}
\end{minipage}
\caption{ (Color online). 
Top: The normalized likelihood $L_{\textrm{SN}}$ for the supernova as a function of  $\tilde{\Omega}_{\Lambda}$.
Bottom: Contours of the normalized likelihood $L$ for both the radiation temperature and the growth rate. 
To calculate $L_{\textrm{SN}}$, the \textit{Union 2.1} set of $580$ type Ia supernovae is used \cite{Suzuki_2012}. 
(For details, see the text.)
The likelihood $L_{\textrm{GR}}$ for the growth rate is from Ref.\ \cite{Koma7}, whereas $L_{\textrm{RT}}$ for the radiation temperature is replotted from Fig.\ \ref{Fig--L_mu_Lambda-T}.
For clarity, the contours of $L =0.2$ and $0.8$ are plotted in the bottom panel.  }
\label{Fig--L_mu_Lambda-T_fc}
\end{figure}

Figure\ \ref{Fig--L_mu_Lambda-T} plots the contours of the normalized likelihood $L_{\textrm{RT}}$ in the $(\tilde{\Omega}_{\Lambda}, \tilde{\mu})$ plane. 
In this figure, $(\tilde{\Omega}_{\Lambda}, \tilde{\mu}) = (\Omega_{\Lambda}, 0)$ corresponds to a pure $\Lambda$CDM model for $\Omega_{\Lambda}$.
Consider the contours of $L_{\textrm{RT}}$ for $\tilde{\Omega}_{\Lambda} \approx 0.7$.
It might be seen that low dissipation ($0 \leq \tilde{\mu} \lessapprox 0.01$) corresponds to high $L_{\textrm{RT}}$ regions when $\tilde{\Omega}_{\Lambda} \approx 0.7$.
That implies that a weakly dissipative Universe is compatible with the current data.
A similar result has been discussed previously \cite{Koma7}, in which the growth rate for clustering (related to structure formations) was examined.  
Accordingly, a weakly dissipative universe is likely proposed.
This expectation may be confirmed by the bottom panel of Fig.\ \ref{Fig--L_mu_Lambda-T_fc} which shows the contours of the normalized likelihood $L$ not only for the radiation temperature but also for the growth rate.
To confirm the expectation in more detail, a joint likelihood analysis is performed later.

In the bottom panel of Fig.\ \ref{Fig--L_mu_Lambda-T_fc}, the normalized likelihood $L_{\textrm{GR}}$ for the growth rate is taken from Ref.\ \cite{Koma7}. 
In that reference, a likelihood analysis for the growth rate in a modified entropic-force model was performed using the same method. 
The modified entropic-force model is equivalent to the modified dissipative model considered here.
Therefore, $L_{\textrm{GR}}$ for the growth rate in the present model is the same as that in Ref.\ \cite{Koma7}.

In addition, using the same method, the normalized likelihood for the distance modulus (related to supernova data points) is discussed. 
In the present study, we call this the normalized likelihood $L_{\textrm{SN}}$ for the supernova.
For the supernova data points, the \textit{Union 2.1} set of $580$ type Ia supernovae (up to redshift $z= 1.414$) is used \cite{Suzuki_2012}. 
The chi-squared function is 
\begin{equation}
\chi_{\textrm{SN}}^{2} (\tilde{\Omega}_{\Lambda}, H_{0})
= \sum\limits_{i=1}^{580} { \left[ \frac{   
                                                         \mu_{\textrm{obs}} (z_{i} ) - \mu_{\textrm{cal}} (z_{i}, \tilde{\Omega}_{\Lambda}, H_{0} )  }{ \sigma_{i}^{\textrm{SN}} }   
                                     \right]^{2}   }   
\label{eq:chi_mu_SN}
\end{equation}
and the distance modulus $\mu$ is defined as
\begin{equation}
 \mu = 5 \log d_{L} + 25   
\end{equation}
where the luminosity distance $d_{L}$ is from Eq.\ (\ref{eq:dL_00}).
Keep in mind that the distance modulus $\mu$ is not the dissipation rate $\tilde{\mu}$. 
In this analysis, $\tilde{\Omega}_{\Lambda}$ and $H_{0}$ are treated as free parameters and sampled in steps of $0.001$ and $0.02$, respectively.
That is, the normalized likelihood $L_{\textrm{SN}}$ does not depend on the dissipation rate $\tilde{\mu}$.
Consequently, $L_{\textrm{SN}}$ in the $(\tilde{\Omega}_{\Lambda}, H_{0})$ plane is obtained.
Using $L_{\textrm{SN}}(\tilde{\Omega}_{\Lambda}, H_{0})$, the maximum value at each $\tilde{\Omega}_{\Lambda}$ can be determined.
The value is plotted as a function of $\tilde{\Omega}_{\Lambda}$ in the top panel of Fig.\ \ref{Fig--L_mu_Lambda-T_fc}.
In this panel, the maximum value is obtained at $\tilde{\Omega}_{\Lambda} = 0.722^{+0.030}_{-0.031}$ (with $H_{0} = 70.04^{+0.40}_{-0.38}$).
This result is consistent with that examined in Ref.\ \cite{Suzuki_2012}.

\begin{figure} [t] 
\begin{minipage}{0.495\textwidth}
\begin{center}
\scalebox{0.33}{\includegraphics{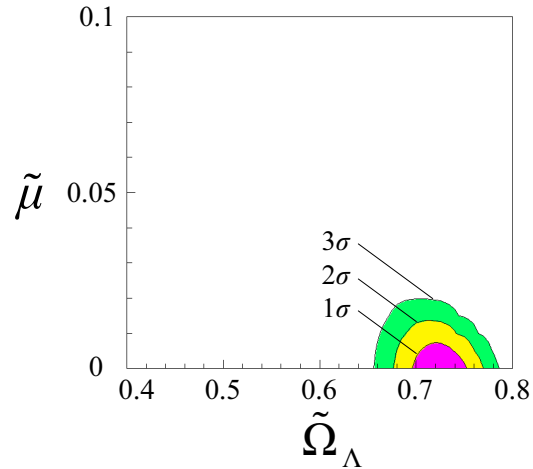}}
\end{center}
\end{minipage}
\caption{ (Color online). Contours of the normalized likelihood $L_\textrm{total}$ for the joint likelihood analysis in the $(\tilde{\Omega}_{\Lambda}, \tilde{\mu})$ plane.
The likelihood function is normalized using the maximum value, obtained for $(\tilde{\Omega}_{\Lambda}, \tilde{\mu}) = (0.725, 0)$.
The contours of $1 \sigma$, $2 \sigma$, and $3 \sigma$ confidence levels (corresponding to $- 2 \ln L_\textrm{total}$ equal to $2.30$, $6.16$, and $11.81$, respectively \cite{Sola_2015c}) are plotted.
That is, the contours correspond to $ L_\textrm{total} = 3.17 \times 10^{-1}$, $4.60 \times 10^{-2}$, and $2.73 \times 10^{-3}$, respectively.  }
\label{Fig--Lt_mu_Lambda-T}
\end{figure}

A weakly dissipative universe likely describes the observed data, as mentioned previously.
To confirm this expectation, a joint likelihood analysis is performed, using the three normalized likelihood functions.
For the joint likelihood analysis, a combined likelihood function $L_{\textrm{total}}$ is defined by
\begin{equation}
L_{\textrm{total}} = L_{\textrm{RT}} \times L_{\textrm{GR}} \times L_{\textrm{SN}}  .
\label{eq:L-joint}
\end{equation}
In this analysis, $\tilde{\Omega}_{\Lambda}$ and $\tilde{\mu}$ are sampled in steps of $0.005$.
The obtained $L_{\textrm{total}}$ is normalized.
The contours of $L_{\textrm{total}}$, corresponding to $1 \sigma$, $2 \sigma$, and $3 \sigma$ confidence levels, are plotted in Fig.\ \ref{Fig--Lt_mu_Lambda-T}.
The region surrounded by the likelihood contours is close to zero dissipation, i.e., $\tilde{\mu} =0$, when $\tilde{\Omega}_{\Lambda} \approx 0.7$.
In fact, the maximum value of $L_{\textrm{total}}$ is obtained for $(\tilde{\Omega}_{\Lambda}, \tilde{\mu}) = (0.725, 0)$, 
where an upper limit of the dissipation rate is $\tilde{\mu} \approx 0.01$ at the $1 \sigma$ confidence level.
In this sense, a weakly dissipative model proposed here is very similar to the standard $\Lambda$CDM model.

As has been noted in Refs.\ \cite{SZ_1970etc,Chluba_2014}, adiabatic photon production does not affect a blackbody spectrum if radiation fields are not thermalized. 
Also, Chluba has recently examined the influence of a CMB (cosmic microwave background) spectral distortion and adiabatic photon production processes on the $T$--$z$ relation in detail \cite{Chluba_2014}.
Consequently, it is found that the photon production process does not affect a blackbody spectrum except when the process has a very special energy dependence.
In contrast, the CMB spectral distortion affects the CMB temperature.
The CMB distortion can be constrained using COBE/FIRAS limits. 
(For details, see Ref.\ \cite{Chluba_2014}.)
Accordingly, if adiabatic photon production for CCDM models \cite{Lima_2000} is assumed, 
a weakly dissipative model discussed here should be further constrained because the CMB distortion is neglected in this study.
That is, the dissipation in the universe is expected to be smaller than the above mentioned one especially when the radiation temperature is discussed.
Similarly, the recent data \cite{Martino_2015,Luzzi_2015,Planck2015} imply a more weakly dissipative model. 
For example, the simple $T$--$z$ relation of the form $T = T_{0}(1+z)^{1-\beta} $ has been investigated in the Planck 2015 results \cite{Planck2015}.
In that reference, one finds $\beta = (0.2 \pm 1.4) \times 10^{-3}$, where a recombination redshift of $z=1100$ is adapted.
Substituting $z =1100$, $\beta = (0.2 \pm 1.4) \times 10^{-3}$, and $T_{0} = 2.725 \pm 0.002$ K (from Table\ \ref{tab-T-z_obs}) into the simple $T$--$z$ relation, we have $T=2965 \sim 3028$ K.
Therefore, when $z =1100$, upper limits of the dissipation rate can be estimated from this result, using Eq.\ (\ref{eq:dotT_T_radiation_T-a_z_2}).
The upper limit is approximately $\tilde{\mu} = 0.018$ \cite{C100}.
This constraint is consistent with the confidence levels shown in Fig.\ \ref{Fig--Lt_mu_Lambda-T}.

\section{Conclusions}
\label{Conclusions}

The radiation temperature--redshift relation has been examined in a dissipative universe.
A phenomenological modified dissipative model has been developed that includes two constant terms, assuming a homogeneous, isotropic, and spatially flat universe.
The model is equivalent to an extended $\Lambda$CDM model in a dissipative universe.
Therefore, it behaves as if a nonzero cosmological constant $\Lambda$ and a dissipative process are operative.
A general radiative temperature law \cite{Lima_2000,Lima_1992e,Lima_1992b,Lima_2014b} has been used in the model, to deduce a $T$--$z$ relation for a dissipative universe.
That relation has been computed in the late universe as a function of a dissipation rate ranging from $\tilde{\mu} =0$, corresponding to a nondissipative $\Lambda$CDM model, to $\tilde{\mu} =1$, corresponding to a fully dissipative CCDM model. 

The results confirm that the $T$--$z$ relation for $\tilde{\mu} =0$ obeys a linear law in equilibrium.
However, the calculated radiation temperature $T$ gradually deviates from the linear law with increasing $\tilde{\mu}$, even if the background evolution of the universe is not altered. 
In particular,  $T$ for $\tilde{\mu} =1$ is nonlinear because the effective equation-of-state parameter $w_{e}$ varies with time in a dissipative universe. 
In contrast, $T$ for low $\tilde{\mu}$ agrees with observations when $\tilde{\Omega}_{\Lambda}  = 0.685$ (i.e., the background evolution is equivalent to that of a fine-tuned pure $\Lambda$CDM model).
This agreement indicates that low dissipation describes the radiation temperature--redshift relation.
That is, the dissipation rate is constrained by the observed $T$--$z$ relation, even if density perturbations are not treated.
The present study thus provides new insights into a dissipative universe. 

To examine the influence of $\tilde{\Omega}_{\Lambda}$, a likelihood analysis has been performed. 
A low-$\tilde{\mu}$ high-$\tilde{\Omega}_{\Lambda}$ universe and a high-$\tilde{\mu}$ low-$\tilde{\Omega}_{\Lambda}$ universe have high likelihoods, consistent with previous work \cite{Koma7} in which a growth rate for clustering (related to structure formations) was examined.  
However, higher and lower $\tilde{\Omega}_{\Lambda}$ values are inconsistent with supernova data. 
Accordingly, a weakly dissipative universe ($\tilde{\mu} \lessapprox 0.01$) for $\tilde{\Omega}_{\Lambda}  \approx 0.7$ is a viable scenario.
To examine this scenario in more detail, a joint likelihood analysis has been performed.
Consequently, a weakly dissipative model proposed here is found to be very similar to the standard $\Lambda$CDM model because the expected dissipation rate is small. 
Interestingly, recent works of the radiation temperature \cite{Planck2015,Martino_2015,Luzzi_2015,Chluba_2014} imply a more weakly dissipative model.
However, the properties of a weakly dissipative model differ from those of a nondissipative $\Lambda$CDM model.
Therefore, further observations are necessary to determine whether a low-dissipation model is valid. 
It should be noted that a fully dissipative CCDM model (for which $\tilde{\mu} =1$) agrees with observations of the growth rate if a negative sound speed \cite{Lima2011} and the existence of clustered matter \cite{Ramos_2014-2014b} are assumed.

\begin{acknowledgements}
The authors wish to thank Professors J. Chluba, G. Luzzi, and I. de Martino for very valuable comments. 
\end{acknowledgements}


\begin{thebibliography}{99}
\bibitem{C100}  

When $z=1100$, it is difficult to compute the radiation temperature from Eq.\ (\ref{eq:dotT_T_radiation_T-a_z_2}) because a value exceeds the range of a double precision representation in the present calculation.
However, $T/(1+z)$ from Eq.\ (\ref{eq:dotT_T_radiation_T-a_z_2}) is confirmed to be constant for $ 8 \leq z \leq 100$.
That is, $T/(1+z)$ for $z=100$ is expected to be the same as that for $z=1100$.
Accordingly, this extrapolation is used to estimate the upper limit of $\tilde{\mu}$ for $z=1100$.
\end{thebibliography}
\end{document}